\documentclass[twocolumn,showpacs,preprintnumbers,amsmath,amssymb,prb]{revtex4-1}

\usepackage[active]{srcltx}
\usepackage{graphicx}
\usepackage{dcolumn}
\usepackage{bm}

\usepackage{hyperref}
\usepackage{color}
\usepackage[a4paper]{anysize}

\definecolor{darkred}{rgb}{0.5,0.0,0.0}
\definecolor{darkgreen}{rgb}{0.0,0.5,0.0}
\definecolor{darkblue}{rgb}{0.0,0.0,0.5}
\hypersetup{colorlinks=true,
citecolor=darkred,
linkcolor=darkgreen,
urlcolor=darkblue}

\newcommand{\im}{\ensuremath{\mathrm{Im}\,}}

\def\gsim{\lower.35em\hbox{$\stackrel{\textstyle>}{\textstyle\sim}$}}
\def\lsim{\lower.35em\hbox{$\stackrel{\textstyle<}{\textstyle\sim}$}}
\begin{document}

\title{Spin-charge separation of plasmonic excitations in thin topological insulators}

\author{T. Stauber and G. G\'omez-Santos}
\affiliation{Departamento de F\'{\i}sica de la Materia Condensada, Instituto Nicol\'as Cabrera and Condensed Matter Physics Center (IFIMAC), Universidad Aut\'onoma de Madrid, E-28049 Madrid, Spain}
\author{L. Brey}
\affiliation{Departamento de Teor\'{\i}a y Simulaci\'on de Materiales, Instituto de Ciencia de Materiales de Madrid, CSIC, 28049 Cantoblanco, Spain}

\begin{abstract}
We discuss plasmonic excitations in a thin slab of a topological insulators. In the limit of no hybridization of the surface states and same electronic density of the two layers, the electrostatic coupling between the top and bottom layers leads to optical and acoustic plasmons which are purely charge and spin collective oscillations. We then argue that a recent experiment on the plasmonic excitations of Bi${}_2$Se${}_3$ [Di Pietro et al, Nat. Nanotechnol. {\bf 8}, 556 (2013)] must be explained by including the charge response of the two-dimensional electron gas of the depletion layer underneath the two surfaces. We also present an analytic formula to fit their data.
\end{abstract}



\maketitle

\section{Introduction}
Typical 3D topological insulators (TI) like Bi${}_2$Se${}_3$ or Bi${}_2$Te${}_3$ are layered materials with repeating unit cells of hexagonal structure consisting of 5 layers. Due to the strong spin-orbit coupling, they display protected surface states that are characterized by a single Dirac cone whereas the bulk states show a full insulating gap.\cite{Hsieh09,Chen09,Xia09} Dirac carriers at the surface of a TI remind on graphene, but whereas in graphene, it is momentum and pseudo-spin that are correlated giving rise to phenomenons such as Klein tunneling, it is momentum and real spin which are locked in the case of these topologically protected edge states.\cite{Hasan10,Qi11} 

The collective modes of this "helical metal" were first discussed by Zhang and co-workers focusing on the curious fact that density fluctuations induce transverse spin fluctuations and vice versa. A transverse spin wave can be generated by a transient spin grating consisting of two orthogonally polarized non collinear incident beams.\cite{Raghu10} To detect the induced charge density wave, one measures e.g. the spatial modulation of reflectivity. These spin-plasmons were also discussed in terms of the plasmon wave function.\cite{Efimkin12,Efimkin12b}  

The charge response of a helical metal is identical to the charge response of graphene apart from a factor of $4=g_vg_s$ where $g_s$ and $g_v$ are the spin and valley degeneracy in graphene. The plasmonic excitations are thus just given by graphene plasmons first discussed in Refs. \onlinecite{Wunsch06,Hwang07} with the dispersion relation $\omega\sim\sqrt{q}$. Here, we want to ask if there are other signatures of the helical metal apart from the possibility of exciting them via transverse spin waves.

Dirac cones must come in multiples of two and the single Dirac cone on one surface naturally finds its pair on the opposite side. For slab thicknesses $d$ larger than six quintuple layers, i.e., $d\approx6$nm, the surface wave functions on opposite sides do not overlap and can thus only couple electrostatically.\cite{Linder09,Liu10,Lu10,Eremeev10} For thick samples and large wave numbers, $qd\gg1$, this coupling can safely be neglected, but for $qd<1$, there will be changes in the plasmon dispersion due to the emergence of a optical (in-phase) and acoustic (out-of-phase) mode. In the context of topological insulators, this was first discussed in Ref. \onlinecite{Profumo12} and also in Ref. \onlinecite{Stauber12}.

\begin{figure}
\centering
  \includegraphics[width=\columnwidth]{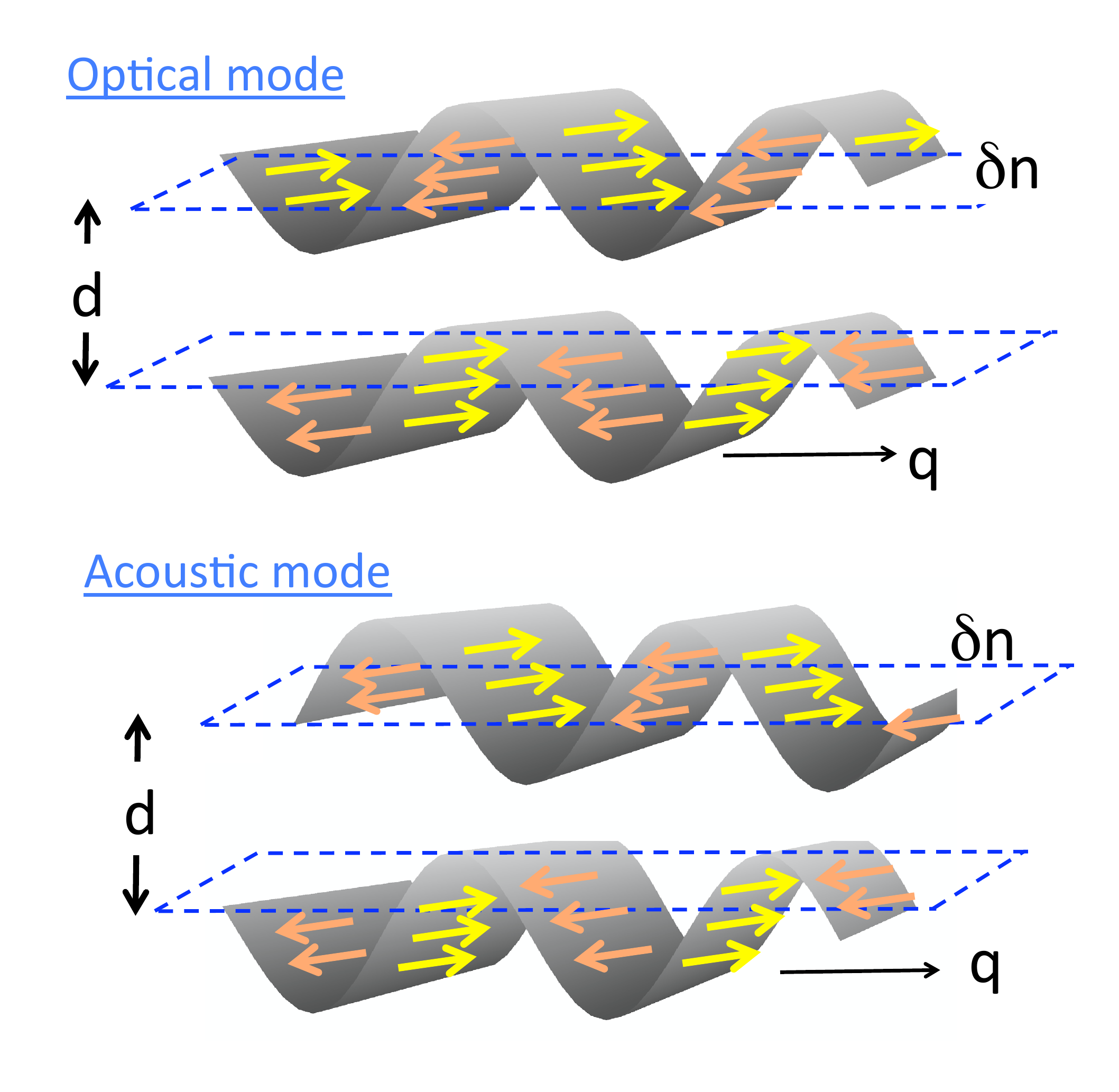}
  \caption{(color online):  Schematic picture of the spin-charge separation. For the optical mode (upper panel), the charge waves are in-phase and the spin waves are in different directions for the top and bottom layer. This leads to an effective (pure) charge wave. For the acoustic mode (lower panel), the charge modulations are opposite for the top and bottom layer and the spin waves point in the same direction for the top and bottom layer. This leads to an effective (pure) spin wave.}
  \label{SpinChargeSep}
\end{figure}

A thin topological insulator slab thus seems to mimic double-layer graphene which were recently experimentally fabricated.\cite{Kim11,Britnell12} But the Dirac cone on one TI surface is {\it not} an identical copy of the Dirac cone on the other surface because the sign of the Fermi velocity must be opposite for the two Dirac cones.\cite{Silvestrov12} This means that the spin locked to the charge momentum is polarized in opposite directions on the two sides which has the curious consequence that in-phase and out-of-phase oscillations can be purely charge- and spin-like, respectively, see Fig. \ref{SpinChargeSep}. Optical plasmon excitations have recently been detected using infrared spectroscopy,\cite{DiPietro13} but no comment was made on their possible pure charge-like character. 

The objectives of this paper are the study of the dispersion and the spin and charge character of the optical and acoustic modes of a thin slab of a TI. We also compare our results with the available experimental data. One of our main conclusions is that in order to explain the experimental data presented in Ref. \onlinecite{DiPietro13}, it is necessary to include, in addition to the response of the Dirac carriers, the charge response of the two-dimensional electron gas (2DEG) of the depletion layer underneath the TI surface.

The paper is organized as follows. We first present analytical results explicitly showing the spin-charge separation of the in-phase and out-of-phase mode in the case of equal charge density on the top and bottom layer. We then solve the plasmon dispersion numerically and analyze the spin-charge separation for general wave numbers. In Sec. III,  we compare our results with the available experimental data and close with conclusions.

\section{Plasmons of a thin TI slab}
The Hamiltonians describing the top and bottom surfaces of a topological insulators slab are
\begin{equation}
H^{T/B} = \pm\hbar v_F  \left( \begin{array}{cc}
0 & -k_x -  i k_y  \\
-k_x +i k_y  & 0  
\end{array} \right) \;.
\end{equation} 
Note that the difference between the top and the bottom Hamiltonians is just the sign of the Fermi velocity, $v_F$.

In topological insulators there is a constaint between the response of the charge density, $\rho$, and the response of the transverse spin density, $S_{\perp}$.\cite{Raghu10,Efimkin12,Efimkin12b}
In the case of a TI film, the tensor density and transverse spin susceptibility at low energy takes the form, 
\begin{equation}
\chi _0 (q,\omega) =  \left( \begin{array}{cc}
\chi _0 ^T  & 0  \\
0  & \chi _0^B  
\end{array} \right)
\end{equation}
with 
\begin{equation}
\chi _0 ^{T/B} (q,\omega)  =   \left( \begin{array}{cc}
1 & \pm x  \\
\mp x  & -x^2  
\end{array} \right) \chi _{\rho\rho}^{T/B} (q,\omega)
\label{chi0}
\end{equation}
where $x= \frac  {\omega}{q v_F}$ and $\chi _{\rho\rho}^{T/B} (q,\omega)$ the (scalar) density-density correlation function of a single Dirac cone corresponding to the top/bottom surface. In the relevant region for plasmons ($\omega>v_Fq$) and in the absence of dissipation ($\im\chi _{\rho\rho}^{T/B}=0$), $\chi _{\rho\rho}^{T/B}$ is given by\cite{Wunsch06,Hwang07} 
\begin{eqnarray}
\label{FullResponse}
\chi_{\rho\rho}^{T/B}&=&-\frac{\mu_{T/B}}{2\pi (\hbar v_F)^2}+\frac{1}{16\hbar\pi}\frac{q^2}{\sqrt{\omega^2-(v_Fq)^2}}\\\nonumber
&\times&\left[G\left(\frac{2\mu_{T/B}+\hbar\omega}{\hbar v_Fq}\right)-G\left(\frac{2\mu_{T/B}-\hbar\omega}{\hbar v_Fq}\right)\right]\;,
\end{eqnarray}
with $G(x)=x\sqrt{x^2-1}-\cosh^{-1}(x)$ for $x>1$ and $\mu_{T/B}$ are the chemical potentials of the top/bottom surface, respectively. The non-diagonal terms in Eq. (\ref{chi0}) appear because of the topological spin-charge coupling. These terms change sign for $\chi _0 ^T$ and $\chi _0 ^B$ because velocities on opposite surfaces have opposite signs. 

Introducing the long-ranged Coulomb interaction within the random-phase approximation (RPA), only the charge sectors of the opposite surfaces will be coupled and the system factorizes. But in order to make the spin-charge separation more explicit, we will keep the full tensor structure. The RPA susceptibility is given by
\begin{equation}
\label{ResponseFunction}
\chi^{RPA}= \chi _0 (q,\omega)\left [ 1-v(q)\chi _0 (q,\omega) \right ] ^{-1}
\end{equation}
with 
\begin{equation}
v(q)=\left(\begin{array}{cccc}
v_T(q) & 0 & v_{TB}(q)& 0  \\
0  & 0 & 0& 0 \\
v_{TB}(q)& 0 & v_B(q)& 0 \\
0&0&0&  0
\end{array}    \right )
\end{equation}
where $v_{T/B}(q)$ and $v_{TB}(q)$ are the intra- and interlayer Coulomb interaction, respectively. Above, we have neglected the finite width of the TI surface states which are spread over the upper(lower)most quintuple layer. Still, it serves as a first approximation and compared to the experimental slab width, the spread of the surface state is about two orders of magnitudes smaller which justifies our approach.

For different dielectric media on the top ($\epsilon_T$) , center ($\epsilon_{TI}$) and bottom ($\epsilon_B$), the general expressions for the intra- and interlayer are given by\cite{Profumo12,Badalyan12,Scharf12,Stauber12} $v_{T/B}=[\cosh(qd)+(\epsilon_{B/T}/\epsilon_{TI})\sinh(qd)]v_{TB}(q)$ and $v_{TB}=e^2\epsilon_{TI}/(\varepsilon_0qN)$  with $N=\epsilon_{TI}(\epsilon_T+\epsilon_B)\cosh(qd)+(\epsilon_T\epsilon_B+\epsilon_{TI}^2)\sinh(qd)$. The width of the topological insulator is denoted by $d$ and it is obvious that the small parameter is given by $qd$, i.e., there will always be hybridization for small enough $q$.\footnote{We neglect retardation effects. The general formulas can be found Ref. \onlinecite{Stauber12}.}

 The collective excitations are determined by the zeros of $\det(1-v(q)\chi _0)$,
\begin{equation}
\label{det}
(1-v_T\chi_{\rho\rho} ^T)( 1-v_B\chi_{\rho\rho} ^B) - v _{TB} ^2 \chi_{\rho\rho} ^T \chi_{\rho\rho} ^B=0\;.
\end{equation}
For the full response function, Eq. (\ref{FullResponse}), the above equation needs to be solved numerically. But we will show that many results relevant for experiments can also be obtained analytically.

\subsection{Analytical results}
To proceed analytically, we use the long-wavelength ($q\to0$) or local approximation of the charge response
\begin{equation}
\label{local}
\chi_{\rho\rho}^{T/B}=\frac{\mu_{T/B}}{4\pi\hbar^2}\frac{q^2}{\omega^2}\;,
\end{equation}
valid for $\omega\gg v_Fq$ and $\mu_{T/B}\gtrsim\hbar\omega$. Assuming also $qd\ll1$, we obtain two modes, the optical (in-phase) and acoustic (out-of-phase) mode 
\begin{eqnarray}
\label{OpticalMode}
\omega _ + ^2 &  = & \frac {\alpha_d v_F(\mu _T + \mu _B )}{\hbar(\epsilon_T+\epsilon_B)} q\;, \\\label{acousticMode}
\omega _- ^2 &  = & \frac {\alpha_d v_F}{ \hbar\epsilon_{TI}} \frac {\mu _T  \mu _B } {\mu _T+ \mu _B } dq^2 \equiv v_s^2q^2\;, 
\end{eqnarray}
where we introduced the fine-structure of a Dirac system $\alpha_d=\frac{e^2}{4\pi\varepsilon_0\hbar v_F}$ and defined the sound velocity in case of the acoustic mode, $v_s$.

\begin{figure}
\centering
  \includegraphics[width=\columnwidth]{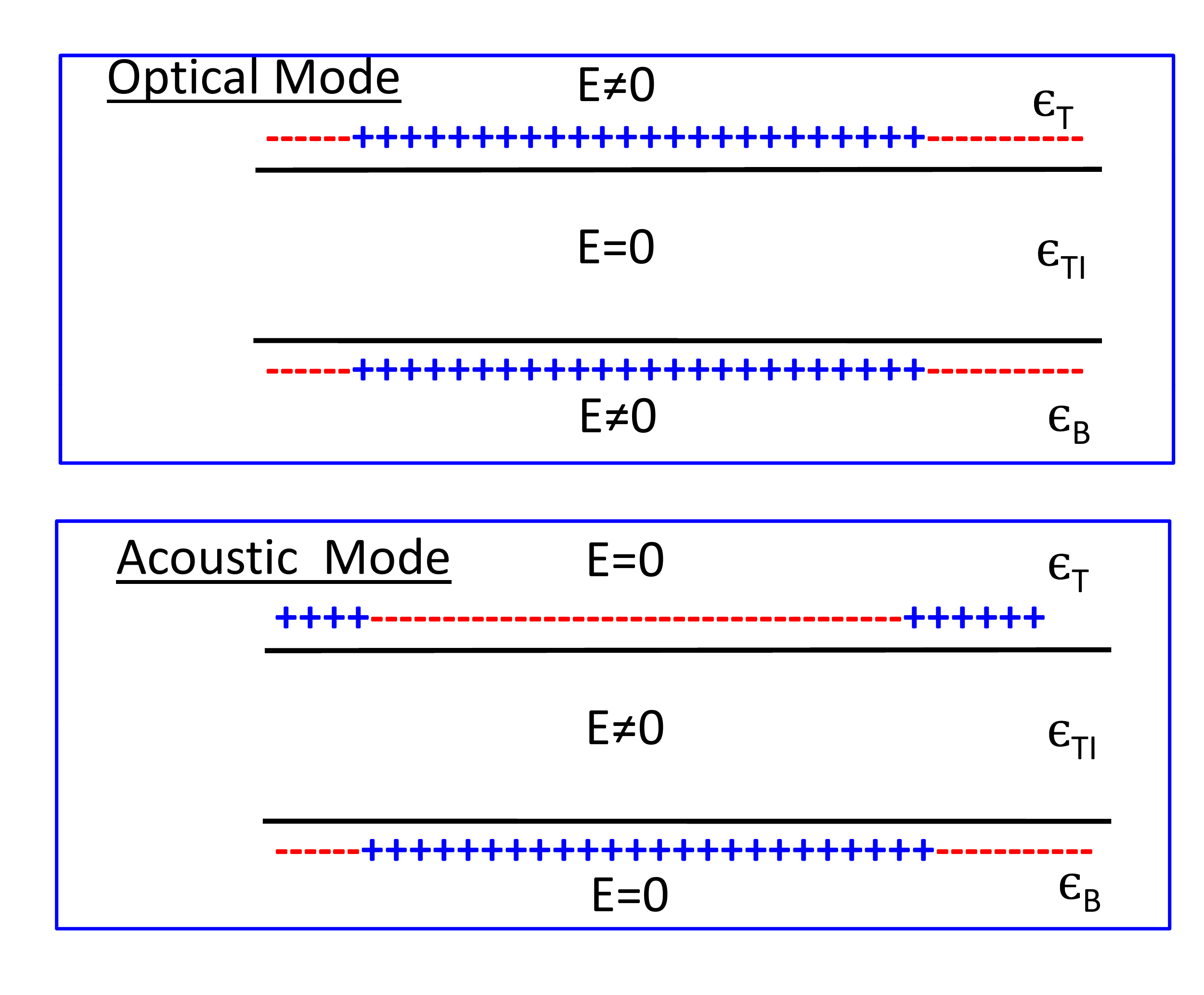}
\caption{(color online):  Schematic picture of the electric field distribution. For the optical mode, there is no electric field inside the TI (upper panel). For the acoustic mode, there is no electric field outside the TI (lower panel).}
  \label{Capacitor}
\end{figure}

Note that the optical mode only depends on the outer dielectric media $\epsilon_T+\epsilon_B$ whereas the acoustic sound velocity only depends on the dielectric medium in the center, $\epsilon_{TI}$. This is generally true since for the optical (in-phase) mode the interfaces have the same homogeneous charge density in the limit $qd\to0$, thus not polarizing the inner medium. For the acoustic (out-of-phase) mode in the same limit, there are opposite homogeneous charge densities on the two sheets just like for a capacitor which in turn does not polarize the surrounding media. This is shown schematically in Fig. \ref{Capacitor}.

The expression for the out-of-phase mode, Eq. \ref{acousticMode}, is only valid for $(k_F^T+k_F^B)d/\epsilon_{TI}\gg1$, $k_F^{T/B}=\mu_{T/B}/v_F$ being the Fermi wave number of the top and bottom surfaces. For the general case, the acoustic mode must be obtained in terms of a Laurent-Taylor expansion including the full expression of the response function $\chi_{\rho\rho}(q,\omega)$.\cite{Santoro88} The square root singularity of $\chi_{\rho\rho}$ at $\omega=v_Fq$ then guarantees that $v_s>v_F$.\cite{Stauber12} The general analytical expression has been obtained in Ref. \onlinecite{Profumo12} and for a system with equal density at the top and at the bottom, $k_F^T=k_F^B\equiv k_F$,  this reads
\begin{equation}
\label{AcousticModeTwo}
v_s=\frac{1+\alpha_dk_Fd/\epsilon_{TI}}{\sqrt{1+2\alpha_dk_Fd/\epsilon_{TI}}}v_F\;.
\end{equation}
In the case $k_Fd/\epsilon_{TI}\ll1$, we can approximate $v_s=v_F$ such that in this limit the spin-like excitations are merging into the particle-hole continuum and the spectral weight of this mode is strongly suppressed.\cite{Stauber12} But for large TI-widths and chemical potential, we will find the sound velocity well inside the Pauli-blocked (protected) region.

For a TI insulator, we have $\epsilon_{TI}\gg\epsilon_T,\epsilon_B$ and the expression for the in-phase mode, Eq. (\ref{OpticalMode}), is only valid for small wave numbers with $qd\epsilon_{TI}/(\epsilon_T+\epsilon_B)\ll1$. For larger $q$, Eq. (\ref{det}) must be replaced by the following compact equation valid for $qd\ll1$ and $\mu_T=\mu_B\equiv\mu$:\footnote{The general expression shall be published elsewhere.}
\begin{equation}
\label{OpticalModeTwo}
\omega _ + ^2= \frac {2\alpha_d v_F\mu q}{\hbar(\epsilon_T+\epsilon_B)}\left[1+\frac{qd\epsilon_{TI}}{\epsilon_T+\epsilon_B}\right]^{-1}
\end{equation}
The optical mode thus approaches a plateau of constant energy for larger $q$ with $\omega_+^2\to 2\alpha_d v_F\mu/(\hbar d\epsilon_{TI})$. Note that this energy only depends on the dielectric constant of the central region, $\epsilon_{TI}$. It is Eq. (\ref{OpticalModeTwo}) that needs to be compared to typical experiments for which $qd\lesssim0.2$, but  $\epsilon_{TI}/(\epsilon_T+\epsilon_B)\approx10$.

\subsection{Charge and Spin density amplitudes of the spin-plasmon collective excitation.}

In a self-sustained excitation the induced potential, in the limit of vanishing external perturbation, has the form, $V^{ind} =v(q)\chi _0 (q,\omega) V^{ind}$.
This implies, again, that the collectives modes correspond to the zeros of $\det(1-v(q)\chi _0)$.
The induced potential is thus obtained from 
\begin{equation}
(1-v(q)\chi _0)_{\omega =\omega_{\pm}} V^{ind}=0\;.
\end{equation}
This yields the typical bonding (+) and and anti-bonding (-) modes:
\begin{equation}
V_+ ^{ind} = V_0\left (   \begin{array}{c}
1\\
0 \\
1\\
 0
\end{array}    \right  )\, \, \, {\rm and} \, \, \,
V _-^{ind} = V_0\left (   \begin{array}{c}
1\\
0 \\
-\frac {\mu _T} {\mu _B}\\
 0
\end{array}    \right  )
\end{equation}
From the induced potentials, we can obtain the charge and spin densities through the relation,
$\rho= \chi _0 (q,\omega)  V^{ind}$.

For the optical mode, $\omega _+ $, we get in the limit $q\to0$
\begin{equation}
\label{OpticalCharge}
\left (   \begin{array}{c}
\rho ^T  \\
S _{\perp} ^T \\
\rho ^B\\
  S _{\perp} ^B
\end{array}    \right  ) =\rho_T\left (   \begin{array}{c}
  1\\
-\sqrt { \frac {\alpha_d (\mu _T+\mu _B)  } {\hbar v_F q(\epsilon_T+\epsilon_B)}}\\
\frac{\mu_B}{\mu_T}\\
 \sqrt { \frac {\alpha_d (\mu _T+\mu _B)  } {\hbar v_F q(\epsilon_T+\epsilon_B)}}\frac{\mu_B}{\mu_T}
 \end{array}    \right  )\;.
\end{equation}

For the acoustic mode, $\omega _-$, we get
\begin{equation}
\label{AcousticSpin}
\left (   \begin{array}{c}
\rho ^T  \\
S _{\perp} ^T \\
\rho ^B\\
  S _{\perp} ^B
\end{array}    \right  ) =\rho_T\left (   \begin{array}{c}
 1\\
- \sqrt {\frac {\alpha_d d}{\hbar v_F\epsilon_{TI}} \frac {\mu _T \mu _B  } {\mu _T+ \mu _B }}\\
-1\\
- \sqrt {\frac {\alpha_d d}{\hbar v_F\epsilon_{TI}} \frac {\mu _T \mu _B  } {\mu _T+ \mu _B }}
 \end{array}    \right  )\;.
\end{equation}

In general, the wavelength of the exciting light $\lambda$ is far greater than the slab thickness $d$ such that the effective charge and spin densities are given by the sum of the two surfaces, $\rho_c=\rho^T+\rho^B$ and $\rho_s=S_\perp^T+S_\perp^B$. This proves the above statement that the collective excitations of thin homogeneously charged TI slabs are purely charge- or spin-like, see Fig. \ref{SpinChargeSep}. Still, they are not comparable with spin collective modes of a clean two-dimensional electron gas with Rashba spin-orbit coupling.\cite{Pletyukhov07} For larger TI slabs, the energy difference between optical and acoustic mode vanishes and the usual spin-plasmon of one layer is recovered. 

We finally note that, for the optical mode, the potential $V$ of the two layers is always in-phase independent of the relative chemical potential. For the acoustic mode, it is the spin-density which is always in-phase independent of the relative chemical potential due to perfect screening of the corresponding charge density. This means that the acoustic mode will show spin-like behavior independent of $\mu_{T/B}$, whereas the  optical mode will loose its pure charge character for unbalanced charge densities $\mu_T\neq\mu_B$.

\subsection{Numerical results}
We will now solve Eq. (\ref{det}) numerically, using the full response function of Eq. (\ref{FullResponse}). We choose $v_F=6\times10^5$m/s and the chemical potential is obtained from the single layer density of $n_{T/B}=1.5\times10^{13}$cm${}^{-2}$.\cite{Bansal12,DiPietro13}. We further set $\epsilon_B=10$, $\epsilon_T=1$ and $\epsilon_{TI}=100$.

\begin{figure}
\centering
  \includegraphics[width=0.99\columnwidth]{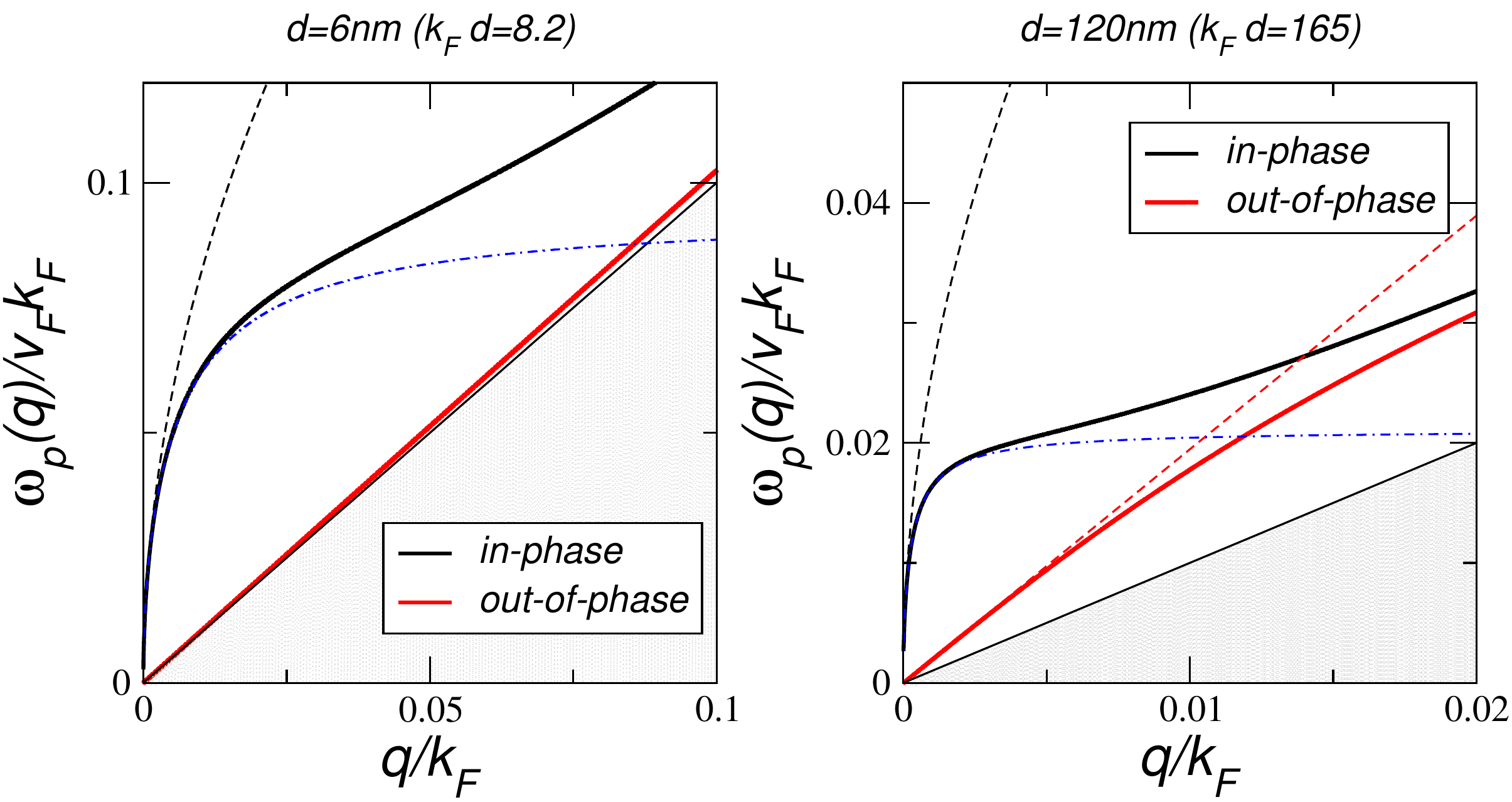}
\caption{(color online): In-phase (black) and out-of-phase (red) plasmonic mode compared to the long wavelength solutions of Eq. (\ref{OpticalMode}) (black dashed), (\ref{AcousticModeTwo}) (red dashed) and Eq. (\ref{OpticalModeTwo}) (blue dotted-dashed) for TI-slab width $d=6$nm (left) and $d=120$nm (right). Also shown the region of intraband excitations bounded by $\omega=v_Fq$ (grey).}
  \label{figure1}
\end{figure}

In Fig. \ref{figure1}, we plot  the in-phase (black) and out-of-phase (red) plasmonic mode for two different slab widths $d=6$nm ($k_Fd=8.2$) (left) and $d=120$nm ($k_Fd=165$). These curves are compared to the analytical formulas for the optical mode of Eq. (\ref{OpticalMode}) (black dashed line) and acoustic mode of Eq. (\ref{AcousticModeTwo}) (red dashed line). Also shown is the analytical formula of Eq. (\ref{OpticalModeTwo}) as blue dotted-dashed line. Finally, the grey region indicates the regime of intraband excitations bounded by $\omega=v_Fq$ .

For a small slab width with $d=6$nm, the acoustic mode is pinned to the electron-hole continuum and Eq. (\ref{AcousticModeTwo}) is almost indistinguishable from the exact solution. But for $d=120$nm, it is already well-separated from $\omega=v_Fq$ and Eq. (\ref{AcousticModeTwo}) starts to differ from the exact solution for $q/k_F\lesssim0.01$. For the optical mode, Eq. (\ref{OpticalMode}) is valid only for small $q$-vectors, but  Eq. (\ref{OpticalModeTwo}) agrees well for $dq\lesssim0.2$ which represents the upper bound of wave numbers in typical experiments. 

For $qd\approx1$, the top and bottom surfaces are almost decoupled and the optical and acoustic plasmon mode begin to approach the (optical) plasmon mode of the decoupled top and bottom surface, respectively. In the case $\epsilon_T=\epsilon_B$, they would merge to the same plasmon mode.

\begin{figure}
\centering
  \includegraphics[width=0.99\columnwidth]{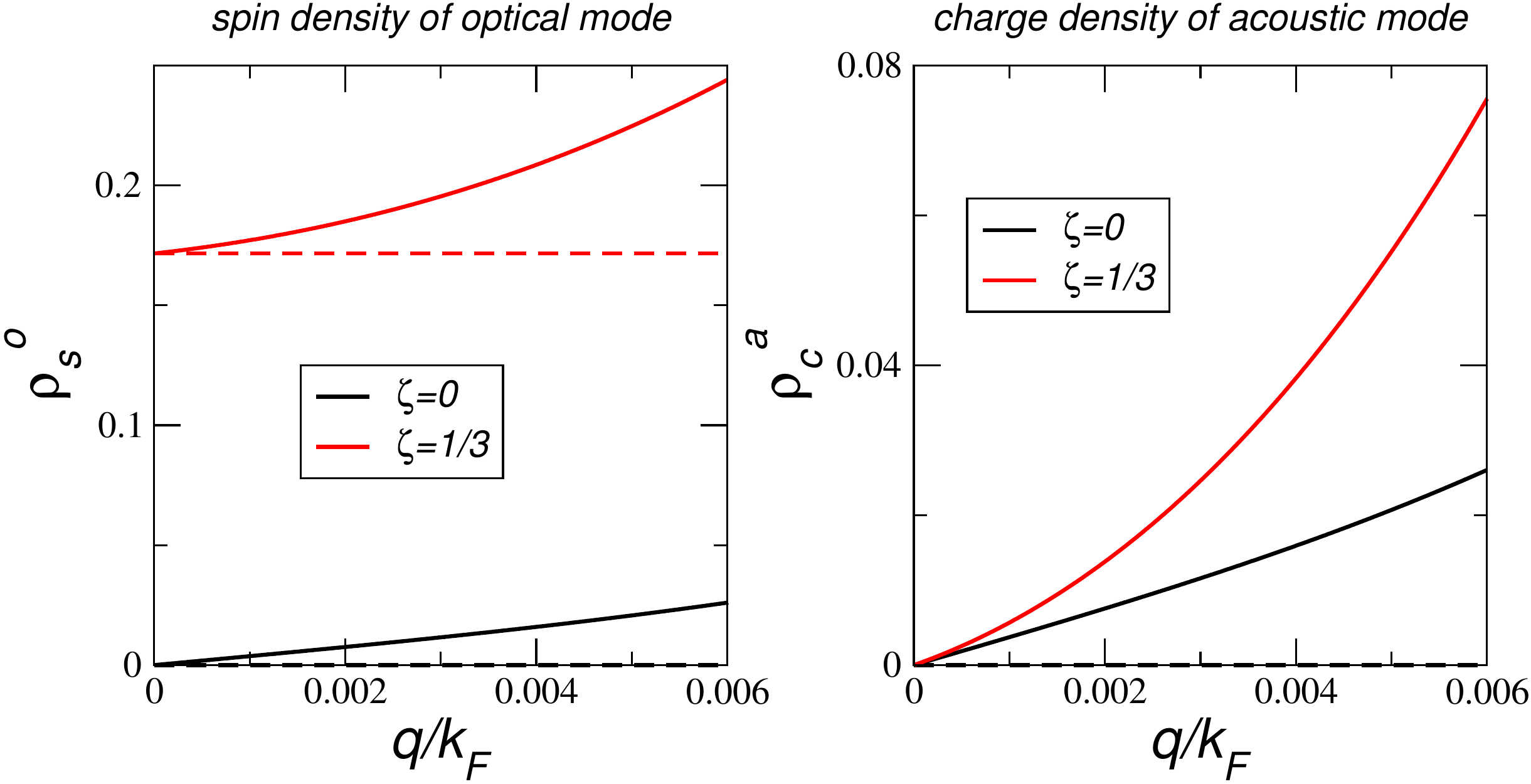}
\caption{(color online): Normalized spin density of the optical mode $\rho_s^o$ (left) and normalized charge density of the acoustic mode $\rho_c^a$ (right) for $\zeta=0$ (black) and $\zeta=1/3$ (red). The slab width is $d=120$nm and the total charge density $n=3\times10^{13}$cm${}^{-2}$. Dashed lines refer to the analytic results of Eqs. (\ref{OpticalCharge}) and (\ref{AcousticSpin}).}
  \label{figureSpinDensity}
\end{figure}

In Fig. \ref{figureSpinDensity}, we analyze in more detail the spin-charge separation beyond the analytical solution of Eqs. (\ref{OpticalCharge}) and (\ref{AcousticSpin}) (full vs. dashed). We plot the total spin density of the optical mode, $\rho_s^o=|S_\perp^T+S_\perp^B|/(|S_\perp^T|+|S_\perp^B|)$, and the total charge density of the acoustic mode, $\rho_c^a=|\rho^T+\rho^B|/(|\rho^T|+|\rho^B|)$. For equal densities on the two surfaces ($\zeta=0$), we find $\rho_s^o=\rho_c^a$ and there remains approximate spin-charge separation up to $qk_F\lesssim0.005$. Wavenumbers in typical experiments are limited by $qk_F\lesssim10^{-3}$, such that spin-charge separation should be observable in this range. For different electronic densities in the top and bottom layer with $\zeta\equiv(n_T-n_B)/(n_T+n_B)=1/3$, the optical mode is accompanied by a suppressed, but finite spin-density. The acoustic mode, nevertheless, remains spin-like for the experimentally relevant regime with $\rho_c^a\lesssim1$\%. Finally, we note that for the symmetric case with $\zeta=0$ and $\epsilon_T=\epsilon_B$, there is perfect spin-charge separation independent of $q$.

\section{Comparison to experimental data}
Due to their large momentum well in the evanescent spectrum, longitudinal plasmons do not couple to propagating electromagnetic radiation. In order to satisfy the conservation of momentum in the photon absorption process, the translational symmetry has to be broken. The necessary momentum is thus provided by patterning the two-dimensional (2D) surface with a sub wavelength periodic structure. For graphene, this has been achieved with a grated geometry showing plasmon resonances with remarkably large oscillator strengths at room temperature.\cite{Ju12,Yan13} It was shown that the absorption spectrum displays the usual Drude peak if the electric field was polarized parallel to the grating whereas it showed a plasmon resonance in the case of perpendicular polarization.

Recently, this strategy was used to measure plasmonic resonances of Bi${}_2$Se${}_3$.\cite{DiPietro13} Here, we will compare the analytical formulas for the optical mode, Eqs. (\ref{OpticalMode}) and (\ref{OpticalModeTwo}),  with the experimental data. Parameters for the Bi${}_2$Se${}_3$ family are used with $v_F=6\times10^5$m/s leading to $\alpha_d=3.7$.\cite{DiPietro13} The chemical potential is obtained from the total electron density of $n=n_T+n_B=3\times10^{13}$cm${}^{-2}$,\cite{Bansal12,DiPietro13} and the density asymmetry of the top and the bottom layer is parametrized by $\zeta=(n_T-n_B)/(n_T+n_B)$. Furthermore, Bi${}_2$Se${}_3$ was grown on a sapphire substrate (Al${}_2$O${}_3$) with bulk dielectric constant $\epsilon_B=10$,\cite{DiPietro13} and we set $\epsilon_T=1$ and $\epsilon_{TI}=100$.\cite{Profumo12} 

\begin{figure}
\centering
  \includegraphics[width=0.99\columnwidth]{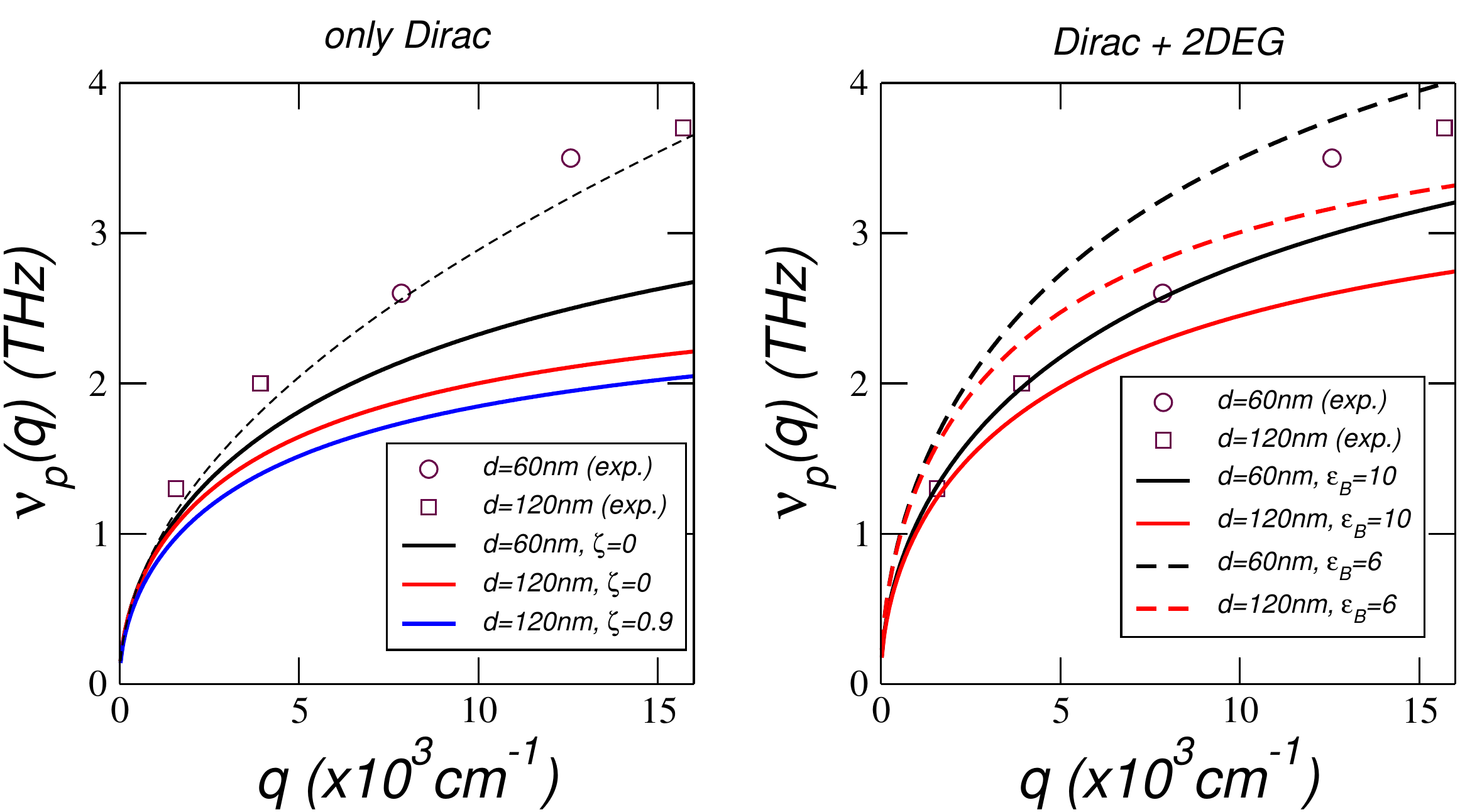}
\caption{(color online): Comparison of the experimental data of Ref. \onlinecite{DiPietro13} (symbols) with the optical mode of Eq. (\ref{OpticalModeTwo}). Right: Including only Dirac Fermions with $\zeta=0$, $d=60$nm (black), $d=120$nm (red) and $\zeta=0.9$, $d=120$nm  (blue). Also shown the optical mode of Eq. (\ref{OpticalMode}) (dashed). Right: Including Dirac Fermions and 2DEG with $\zeta=0$, $d=60$nm (black), $d=120$nm (red) for $\epsilon_B=10$ (full lines) and $\epsilon_B=6$ (dashed lines).}
  \label{Compare}
\end{figure}

On the left hand side of Fig. \ref{Compare}, we plot the in-phase mode of Eq. (\ref{OpticalModeTwo}) for slab widths $d=60$nm (black) and $d=120$nm (red). A more general analytic formula was used to describe the case of asymmetric layer densities ($d=120$nm and $\zeta=0.9$). In the shown regime of small wave numbers $qd\lesssim0.2$, the analytic curves are indistinguishable with the exact numerical solution, but already depend on the slab width $d$. They further lie well below the experimental data which were taken for slab widths $d=60$nm (circles) and $d=120$nm (squares). Let us emphasize that the good agreement with the slab width independent optical mode of Eq. (\ref{OpticalMode}) (dashed line) is pure coincidence.

From the above discussion we conclude that the experimental data cannot be fitted by only assuming Dirac carriers. But a single Dirac cone comprising the topological state can co-exist with a two-dimensional spin-degenerate electron gas (2DEG).\cite{Bianchi10} Accordingly, two conducting channels were identified, one corresponding to the above mentioned Dirac electrons and the other to the 2DEG that develops due to band bending at the two surfaces.\cite{Bansal12} The channels behave independently and the depletion layer extends only within a few nanometers of the surface.\cite{Bansal12} 

Taking the depletion layer into account, changes in number and behavior of plasmon
modes might occur because one now has to consider a response matrix twice as big as the initial one of Eq. (\ref{ResponseFunction}), i.e., a 8x8 instead of a 4x4 matrix. But due to the closeness of the depletion layer to the surface, the only newly emerging modes would be charge-less acoustic-like excitations formed by superpositions of the Dirac carriers and the depletion layer on the same (top or bottom) TI surface. These modes are thus closely pinned to the particle-hole continuum and probably not observable. They further do not affect the modes obtained by the initial 4x4 matrix. 

In the following, we will thus neglect the possibility of particle exchange between the Dirac carriers and depletion layer and also the emerging charge-less acoustic modes. The total charge response can then be taken as the sum of the charge response of the Dirac system and the 2DEG, $\chi_{\rho\rho}^{total}=\chi_{\rho\rho}^{T/B}+\chi_{\rho\rho}^{2DEG}$. The full density response of a 2DEG was derived by Stern,\cite{Stern67} but to describe typical experiments, the local approximation is sufficient which reads
\begin{equation}
\chi_{\rho\rho}^{2DEG}=\frac{\mu^{2DEG}}{\pi\hbar^2}\frac{q^2}{\omega^2}\;.
\end{equation}

There is a unique Fermi energy for both Dirac and 2DEG fermions, 
$\mu^{2DEG}$  and $\mu^{Dirac}$ give its position with respect to the bottom 2DEG 
bottom band and to the Dirac point, respectively. To describe the above experiment, we can thus use Eq. (\ref{OpticalModeTwo}) with $\mu\to\mu^{Dirac}+4\mu^{2DEG}$. The chemical potential of the Dirac system is obtained from the sheet density leading to $\mu^{Dirac}=542$meV, and not from the slightly curved energy dispersion which would yield the lower value $\mu=450$meV.\cite{Bansal12} With $\mu^{2DEG}=60$meV,\cite{Bansal12} we now obtain a reasonable fit to the experimental data for low wave numbers $q\lesssim10^4$cm${}^{-1}$. This can be seen on the right hand side of Fig. \ref{Compare}, where we plot the resonant plasmon frequencies $\nu_p=\omega_+/2\pi$ for slab widths $d=60$nm (black) and $d=120$nm (red) for a dielectric substrate with $\epsilon_B=10$ (full lines). 

The two high-energy plasmon resonances with $q>10^4$cm${}^{-1}$ cannot be well described by our fit and are blue shifted. This is in contrast with our expectations because the dipole-dipole interaction between the patterned nano wires should lead to an additional red-shift compared to the analytic curves for samples with small periodicities;\cite{Nikitin12,Christensen12}  and this shift can be as large as 20\%.\cite{Strait13}  A possible blue shift could be provided by including the frequency dependence of $\epsilon_B$ which might lead to smaller values $\epsilon_B(\omega)<10$. We, therefore, also show curves with $\epsilon_B=6$ (dashed lines) which value was measured for thin ($15$nm) Al${}_2$O${}_3$-films.\cite{Kim09} Also a decrease of $\epsilon_{TI}$ would lead to a blue shift for larger frequency since the high-frequency dielectric constant has a value of $\epsilon_{TI}^\infty\sim25$.\cite{Stordeur92,Nechaev13}
 
\section{Conclusions}
We have analytically and numerically demonstrated how Dirac plasmons in thin TI samples hybridize and that the different sign of the Fermi velocity of the two Dirac cones leads to the curious fact that the in-phase and out-of-phase collective modes can be purely charge- and spin-like, respectively. The peculiar nature of the collective oscillations of TI might have interesting consequences such that thin nano-wires should behave as helical Luttinger liquids.\cite{Egger10} 

We have also analyzed recent experiments measuring plasmonic resonances in Bi${}_2$Se${}_3$ and argued that it is necessary to include the 2DEG in the depletion layer in order to fit the experimental data. Nevertheless, a proper treatment of the dipole-dipole interaction between the patterned nano wires on top of the TI surface is still missing which is planned for future work.

\begin{acknowledgments} 
We thank C. Tejedor, F. J. Garcia-Vidal, M. Ortolani and F. Peeters for helpful discussions. This work has been supported by FCT under grants
PTDC/FIS/101434/2008; PTDC/FIS/113199/2009 and MIC under grants
FIS2010-21883-C02-02;FIS2012-33521;FIS2012-37549-C05-03. 
\end{acknowledgments}
 
\bibliography{topological} 
 \end{document}